\documentclass[reprint,
 amsmath,amssymb,
 aps,
]{revtex4-2}

\usepackage{graphicx}
\usepackage{dcolumn}
\usepackage{bm}
\usepackage{xcolor}

\begin{document}


\title{Spectrally recycling space-time wave packets}

\author{Layton A. Hall$^{1}$}
\author{Ayman F. Abouraddy$^{1}$}
\affiliation{$^{1}$CREOL, The College of Optics \& Photonics, University of Central~Florida, Orlando, FL 32816, USA}

\begin{abstract}
`Space-time' (ST) wave packets are propagation-invariant pulsed optical beams that travel rigidly in linear media without diffraction or dispersion at a potentially arbitrary group velocity. These unique characteristics are a result of spatio-temporal spectral correlations introduced into the field; specifically, each spatial frequency is associated with a single temporal frequency (or wavelength). Consequently, the spatial and temporal bandwidths of ST wave packets are correlated, so that exploiting an optical source with a large temporal bandwidth or achieving an ultralow group velocity necessitate an exorbitantly large numerical aperture. Here we show that `spectral recycling' can help overcome these challenges. `Recycling' or `reusing' each spatial frequency by associating it with multiple distinct but widely separated temporal frequencies allows one to circumvent the proportionality between the spatial and temporal bandwidths of ST wave packets, which has been one of their permanent characteristics since their inception. We demonstrate experimentally that the propagation invariance, maximum propagation distance, and group velocity of ST wave packets are unaffected by spectral recycling. We also synthesize a ST wave packet with group velocity $c/14.3$ ($c$ is the speed of light in vacuum) with a reasonable numerical aperture made possible by spectral recycling.
\end{abstract}

\maketitle

\section{Introduction}

Since Brittingham discovered in 1983 a propagation-invariant (diffraction-free and dispersion-free) pulsed optical beam \cite{Brittingham83JAP}, a host of other wave packets that share this feature have been identified \cite{Saari97PRL,Salo00PRE,Reivelt03arxiv,Christodoulides04OE,Kiselev07OS,Lahini07PRL,FigueroaBook8,Turunen10PO,FigueroaBook14}. Such pulsed beams have been recently dubbed generically `space-time' (ST) wave packets, because their propagation invariance is undergirded by tight spectral associations between the spatial and temporal degrees of freedom \cite{Kondakci16OE,Parker16OE,Porras17OL,Efremidis17OL,Wong17ACSP1,Wong17ACSP2,PorrasPRA18,Wong20AS}. Specifically, each spatial frequency underlying the spatial profile of a ST wave packet is assigned to a single temporal frequency (or wavelength) underlying its temporal profile \cite{Donnelly93PRSLA,Saari04PRE,Longhi04OE}. Nevertheless, exploiting the unique characteristics predicted theoretically for such propagation-invariant wave packets (such as arbitrary group velocities in free space \cite{Salo01JOA,Alexeev02PRL,Zapata06OL,Valtna07OC,Zamboni08PRA,Bonaretti09OE,Bowlan09OL,Kuntz09PRA}) has been hampered in practice by the lack of a precise methodology for spatio-temporal wave-packet synthesis \cite{Turunen10PO}. Recently, a new strategy based on spatio-temporal spectral-phase modulation \cite{Kondakci17NP,Yessenov19OPN} is now helping realize the full potential of ST wave packets. This technique has enabled demonstrations of arbitrary group velocities in free space \cite{Kondakci19NC,Yessenov19OE}, transparent dielectrics \cite{Bhaduri19Optica,Bhaduri20NP}, planar waveguides \cite{Shiri20NC}, and plasmonic interfaces \cite{Schepler20ACSP}, accelerating wave packets \cite{Yessenov20PRL}, non-accelerating Airy wave packets \cite{Kondakci18PRL}, self-healing \cite{Kondakci18OL}, novel ST Talbot effects \cite{yessenov2020veiled}, and even realizations using incoherent light \cite{Yessenov19Optica,Yessenov2019OL}. 

Nevertheless, a fundamental constraint is imposed on ST wave packets because the tight spatio-temporal spectral correlations entail a \textit{proportionality} between the spatial and temporal bandwidths. Such a proportionality is absent from traditional pulsed optical beams in which the spatial and temporal degrees of freedom are more-or-less independent; that is, the spatial beam profile and the temporal pulse linewidth can be manipulated independently. In general, ultrashort optical pulses need not require large numerical apertures. For ST wave packets, on the other hand, the proportionality between the spatial and temporal bandwidths entails that exploiting a large temporal bandwidth will exact a high price in terms of system design: a large spatial bandwidth -- and thus high numerical aperture -- is necessary. In general, the temporal bandwidth $\Delta\omega$ and the spatial bandwidth $\Delta k_{x}$ (in one dimension) are related through $\Delta\omega\!\propto\!(\Delta k_{x})^{2}$ \cite{Kondakci16OE,Kondakci17NP}. This proportionality is particularly exacerbating for ultraslow ST wave packets (group velocity $\widetilde{v}\!\ll\!c$, where $c$ is the speed of light in vacuum) \cite{Yessenov19OE}. Indeed, this bandwidth restriction has been a permanent feature of ST wave packets since their introduction into optical physics. Although progress has been made in exploiting larger temporal bandwidths by relying on lithographically inscribed phase plates \cite{Kondakci18OE,Bhaduri19OL}, the limit on lithographic spatial resolution is nevertheless quickly reached with increased bandwidths. Consequently, this constraint sets upper limits for the practically utilizable light-source bandwidth and the minimally achievable group velocity in free space.

Here we show that the proportionality between spatial and temporal bandwidths that is intrinsic to all propagation-invariant ST wave packets can be circumscribed \textit{without} negative impact on any of their unique characteristics, thus paving the way to reducing the numerical aperture necessary for accommodating large temporal bandwidths and enabling the synthesis of ultraslow pulses in free space. Our strategy relies on spectral `recycling'; that is, each spatial frequency is recycled (or reused) $N$ times in association with distinct but widely separated temporal frequencies, rather than used only once as is the case in typical ST wave packets. Consequently, the temporal spectrum can be increased -- in principle, indefinitely -- while retaining a \textit{fixed} spatial bandwidth. In this approach, the full temporal bandwidth $\Delta\omega$ is divided into $N$ sub-bandwidths $\Delta\omega_{n}$, $n\!=\!1,2,\cdots,N$, each associated with a \textit{fixed} spatial bandwidth that is reduced by a factor $\sim\sqrt{N}$ with respect to its non-recycled counterpart having the same temporal bandwidth. Although the spatio-temporal profile of the spectrally recycled ST wave packet exhibits an additional discrete structure, it nevertheless retains all the essential characteristics of its non-recycled counterpart: it is propagation-invariant and has the same group velocity. We experimentally verify this strategy using $N$ up to 16, which has enabled the realization of ST wave packets with group velocity down to $\widetilde{v}\!=\!c/14.3$ in free space at the same numerical aperture as that of a non-recycled ST wave packet with $\widetilde{v}\!=\!c/2$. These results indicate the possibility of reaching ultralow group velocities in free space, and exploiting new sources of light in synthesizing ST wave packets and fields, such as supercontinuum laser sources, white light sources, and potentially even attosecond pulses.

\begin{figure*}[t!]
\begin{center}
\includegraphics[width=14cm]{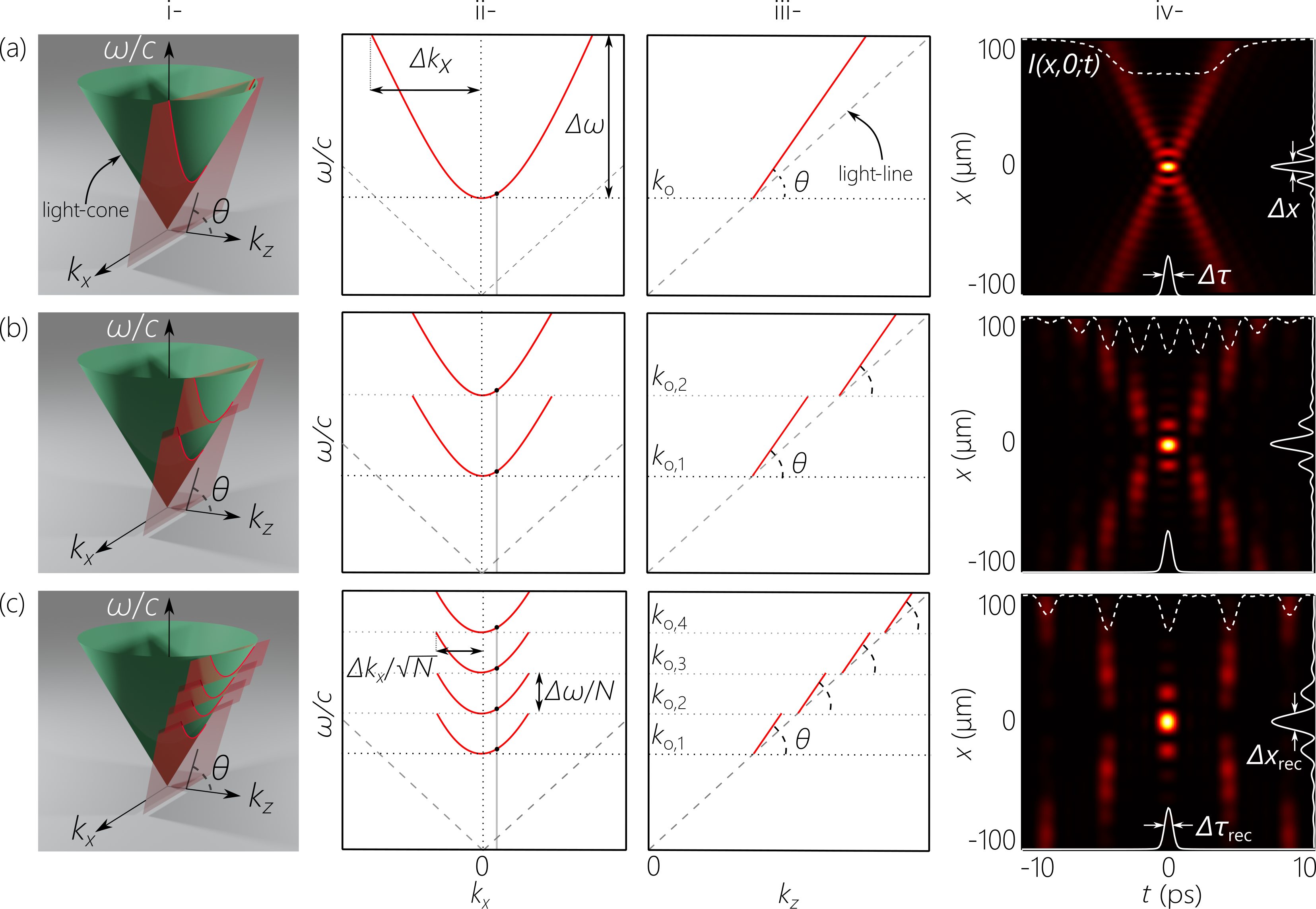} 
\end{center}
\caption{Concept of spectrally recycled ST wave packets. (a) A ST wave packet with a spectral tilt angle $\theta\!=\!60^{\circ}$. (b) A spectrally recycled ST wave packet with $N\!=\!2$ and (c) with $N\!=\!4$. Note that the spectral bandwidth $\Delta k_{x}$ (and thus the numerical aperture of the optical system) is reduced as $N$ increases for the same temporal bandwidth $\Delta\omega$. i- Three-dimensional representation of the spatio-temporal spectral support of the ST wave packet on the surface of the light-cone in $(k_{x},k_{z},\tfrac{\omega}{c})$ space. ii- Spectral projection onto the $(k_{x},\tfrac{\omega}{c})$-plane, and (iii) onto the $(k_{z},\tfrac{\omega}{c})$-plane. Spectral recycling entails that each $k_{x}$ is associated with $N$ distinct temporal frequencies. iv- Spatio-temporal intensity profile $I(x,z;t)$ calculated at a fixed axial plane $z\!=\!0$ using $\lambda_{\mathrm{o}}\!=\!800$~nm and a temporal bandwidth $\Delta\lambda\!=\!2$~nm. We plot along the bottom of the panel in (iv) as a solid white curve the temporal pulse profile at the beam center $I(0,0;t)$ as recorded by a point detector; we plot along the top of the panel as a dashed curve the spatially averaged temporal profile $\int\!dx\,I(x,0;t)$ as recorded by a `bucket' detector; and on the right we plot as a solid white curve the spatial profile at the center of the pulse $I(x,0;0)$. The spatial width $\Delta x$ of $I(x,0;0)$ increases with spectral recycling, but the temporal width $\Delta\tau$ of $I(0,0;t)$ remains fixed.}
\label{fig:Concept}
\end{figure*}

\section{Theory of spectral recycling}

We start our analysis from a traditional optical wave packet $E(x,z;t) = e^{i(k_{\mathrm{o}}z-\omega_{\mathrm{o}}t)}\psi(x,z;t)$ and express its envelope $\psi(x,z;t)$ in terms of an angular spectrum:
\begin{equation}\label{eq:wavepacket}
\psi(x,z;t)=\iint\!dk_{x}d\Omega\,\widetilde{\psi}(k_{x},\Omega)e^{ik_{x}x}e^{i(k_{z}-k_{\mathrm{o}})z}e^{-i\Omega t},
\end{equation}
where $\widetilde{\psi}(k_{x},\Omega)$ is the spatio-temporal spectrum, $\Omega\!=\!\omega-\omega_{\mathrm{o}}$ is the temporal frequency with respect to the carrier frequency $\omega_{\mathrm{o}}$, $k_{\mathrm{o}}\!=\!\tfrac{\omega_{\mathrm{o}}}{c}$, $k_{x}$ (the spatial frequency) is the transverse component and $k_{z}$ is the longitudinal component of the wave vector along the transverse $x$ and axial $z$ coordinates, respectively. For simplicity, we consider the field to be uniform along $y$ ($k_{y}\!=\!0$). The spectral support domain of this pulsed beam is a two-dimensional patch on the surface of the light-cone $k_{x}^{2}+k_{z}^{2}\!=\!(\tfrac{\omega}{c})^{2}$ \cite{Kondakci17NP,Yessenov19OPN}. In contrast, the spatio-temporal spectrum of a propagation-invariant ST wave packet [Fig.~\ref{fig:Concept}(a)] lies at the intersection of the light-cone with the spectral plane
\begin{equation}\label{eq:OmegaKzEquation}
\Omega=(k_{z}-k_{\mathrm{o}})c\tan{\theta},
\end{equation}
where the spectral tilt angle $\theta$ is measured with respect to the $k_{z}$-axis \cite{Kondakci17NP,Yessenov19PRA}; see Fig.~\ref{fig:Concept}(a)-i. Consequently, each spatial frequency $k_{x}$ is associated with a single temporal frequency $\Omega$, $\widetilde{\psi}(k_x,\Omega)\!\rightarrow\!\tilde{\psi}(k_x)\delta(\Omega-\Omega(k_x))$, where $\Omega(k_x)$ is the conic section at the intersection of the spectral plane with the light-cone [Fig.~\ref{fig:Concept}(a)-ii]. The envelope of the ST wave packet takes the form:
\begin{equation}\label{eq:STAngularSpectrum}
\psi(x,z;t)\!=\!\int\!dk_{x}\,\widetilde{\psi}(k_{x}) e^{ik_{x}x}e^{-i\Omega(t-z/\widetilde{v})}\!=\!\psi(x,0;t-z/\tilde{v}),
\end{equation}
which corresponds to a propagation-invariant wave packet that propagates rigidly in free space at a group velocity $\widetilde{v}\!=\!c\tan{\theta}$ \cite{Salo01JOA,Kondakci19NC}. Therefore, tuning $\theta$ changes $\widetilde{v}$ in free space above or below $c$ (without violating relativistic causality \cite{Shaarawi00JPA,SaariPRA18,Saari19PRA,Yessenov19PRA,Yessenov19OE}). According to this construction, higher $\omega$ is associated with lower $k_{x}$ when $\theta\!<\!45^{\circ}$, and with higher $k_{x}$ otherwise. Because the spectral projection onto the $(k_{z},\tfrac{\omega}{c})$-plane in Fig.~\ref{fig:Concept}(a)-iii is a straight line, the wave packet propagates at a well-defined group velocity $\widetilde{v}\!=\!c\tan{\theta}$ without  group velocity dispersion (GVD) \cite{SalehBook07}. Increased deviation of the spectral projection onto the $(k_{z},\tfrac{\omega}{c})$-plane away from the light-line $k_{z}\!=\!\tfrac{\omega}{c}$ [Fig.~\ref{fig:Concept}(a)-iii] is associated with higher $k_{x}$ [Fig.~\ref{fig:Concept}(a)-ii]. 

Finally, the spatio-temporal intensity profile $I(x,z;t)$ at a fixed axial plane $z$ is shown in Fig.~\ref{fig:Concept}(a)-iv, which exhibits a characteristic X-shape \cite{Saari97PRL,Yessenov19OE}. The pulse width $\Delta\tau$ at the beam center, defined as the width of $I(0,z;t)$ [Fig.~\ref{fig:Concept}(a)-iv], is the inverse of the temporal bandwidth $\Delta\omega$. The beam width $\Delta x$ at the pulse center, defined as the width of $I(x,z;0)$, is the inverse of the spatial bandwidth $\Delta k_{x}$.

It is useful to consider the narrowband paraxial limit ($\Delta k_{x}\!\ll\! k_{\mathrm{o}}$ and $\Delta\omega\!\ll\!\omega_{\mathrm{o}}$), whereupon the spectral projection of the conic section onto the $(k_{x},\tfrac{\omega}{c})$-plane can be approximated by a parabola $\tfrac{\Omega(k_x)}{\omega_{\mathrm{o}}}=\tfrac{1}{1-\cot{\theta}}\tfrac{k_x^2}{2k_{\mathrm{o}}^{2}}$ \cite{Bhaduri20NP}. As such, the spatial and temporal bandwidths are related through
\begin{equation}\label{Eq:BandwidthProportion}
\frac{\Delta\omega}{\omega_{\mathrm{o}}}=\frac{1}{|1-\cot{\theta}|}\frac{(\Delta k_{x})^{2}}{2k_{\mathrm{o}}^{2}};
\end{equation}
i.e., $\Delta\omega\!\propto\!(\Delta k_{x})^{2}$. The proportionality constant between $\Delta\omega$ and $(\Delta k_{x})^{2}$ depends sensitively on the spectral tilt angle $\theta$. When $\theta\!\rightarrow\!45^{\circ}$ and $\widetilde{v}\!\rightarrow\!c$, the proportionality constant is large, so that only a small spatial bandwidth $\Delta k_{x}$ is required to accommodate a given temporal bandwidth $\Delta\omega$. However, ultraslow ST wave packets $\widetilde{v}\!\ll\!c$ require $\theta\!\rightarrow\!0^{\circ}$, in which case the proportionality constant is extremely small and a larger $\Delta k_{x}$ is required for the same $\Delta\omega$. For example, with respect to a ST wave packet having $\theta\!=\!44^{\circ}$ ($\widetilde{v}\!=\!0.97c$), reducing the spectral tilt angle to $\theta\!=\!4^{\circ}$ ($\widetilde{v}\!=\!0.07c$) requires increasing $\Delta k_{x}$ by a factor of $\approx\!19$ for the same $\Delta\omega$.

We are now in a position to elucidate our proposed concept of spectral recycling. Spectral recycling aims to \textit{reduce} the separation between the spectral projection of the ST wave packet onto the $(k_{z},\tfrac{\omega}{c})$-plane and the light-line, which \textit{reduces} the maximum $k_{x}$ and hence maintains a low numerical aperture despite increase in the temporal bandwidth. To reduce the spatial bandwidth $\Delta k_{x}$ of the ST wave packet associated with a fixed temporal bandwidth $\Delta\omega$, we `reuse' the spatial frequencies, or `recycle' them. That is, instead of each $k_{x}$ being assigned to a single unique temporal frequency $\omega$, each $k_{x}$ is associated with $N$ distinct and widely separated temporal frequencies: $\omega_{1},\omega_{2},\cdots,\omega_{N}$. Consider a non-recycled ST wave packet of temporal bandwidth $\Delta\omega$ starting at $\omega_{\mathrm{o}}$ with an associated spatial bandwidth $\Delta k_{x}$. Its spectrally recycled counterpart has the same total temporal bandwidth $\Delta\omega$, but is segmented into $N$ sub-bandwidths $\Delta\omega_{n}$ starting at distinct frequencies $\omega_{\mathrm{o},n}$, $n\!=\!1,2,\cdots,N$, which are selected such that the associated spatial bandwidths $(\Delta k_{x})_{n}$ are all equal $(\Delta k_{x})_{n}\!=\!\Delta k_{x}'\!<\!\Delta k_{x}$. That is, the spatial bandwidth of the spectrally recycled ST wave packet is less than that of its non-recycled counterpart.

This concept is illustrated via two examples depicted in Fig.~\ref{fig:Concept}(b,c). In Fig.~\ref{fig:Concept}(b), we show a spectrally recycled ST wave packet with $N\!=\!2$ having the same temporal bandwidth as that of the non-recycled ST wave packet in Fig.~\ref{fig:Concept}(a).  The spatio-temporal spectrum results from the intersection of \textit{two} spectral planes $\omega-\omega_{\mathrm{o},n}\!=\!(k_{z}-k_{\mathrm{o},n})c\tan{\theta}$ with the light-cone [Fig.~\ref{fig:Concept}(b)-i], $k_{\mathrm{o},n}\!=\!\omega_{\mathrm{o},n}/c$, $n\!=\!1,2$, such that the sub-bandwidths are staggered with no spectral gaps [Fig.~\ref{fig:Concept}(b)-ii]. The spatial spectra associated with these two sub-bandwidths in the $(k_{x},\tfrac{\omega}{c})$-plane are both centered at $k_{x}\!=\!0$, and recycling the use of the spatial frequencies results in each $k_{x}$ being now associated with \textit{two} temporal frequencies $\omega$ (one belonging to each sub-bandwidth) -- rather than to one $\omega$ as in the non-recycled ST wave packet. Consequently, the spatial bandwidth of the spectrally recycled ST wave packet $\Delta k_{x}'$ is smaller than that of its non-recycled counterpart despite having the same temporal bandwidth $\Delta\omega$. A further example illustrates the concept of spectral recycling for the case $N\!=\!4$ as shown in Fig.~\ref{fig:Concept}(c). The sub-bandwidths $\Delta\omega_{n}$ and the associated fixed spatial bandwidth $\Delta k_{x}'$ are smaller that those for $N\!=\!1$, and each $k_{x}$ is recycled by associating it with 4 distinct temporal frequencies $\omega$.

In general, holding the recycled spatial spectrum $\Delta k_{x}'$ fixed necessitates selecting unequal sub-bandwidths $\Delta\omega_{n}$ because $\Delta\omega_{n}$ is inversely proportional to $\omega_{\mathrm{o},n}$ for fixed $\Delta k_{x}'$ as seen in Eq.~\ref{Eq:BandwidthProportion}. Consequently, the separations between the distinct temporal frequencies $\omega_{1}$, $\omega_{2}$, ..., $\omega_{N}$ associated with a particular spatial frequency $k_{x}$ are also unequal. Within the paraxial approximation used in Eq.~\ref{Eq:BandwidthProportion}, we have $\omega_{n}-\omega_{n-1}\!=\!(\omega_{\mathrm{o},n}-\omega_{\mathrm{o},n-1})(1-\Omega_{n-1}(k_{x})/\omega_{\mathrm{o},n})$. Only for small bandwidths is $\omega_{n}-\omega_{n-1}\!=\!\omega_{\mathrm{o},n}-\omega_{\mathrm{o},n-1}$. Indeed, in the narrow bandwidth limit $\Delta\omega\!\ll\!\omega_{\mathrm{o}}$, the sub-bandwidths are approximately equal $\Delta\omega_{n}\!\approx\!\tfrac{\Delta\omega}{N}$, and $\Delta k_{x}'\!\approx\!\tfrac{\Delta k_{x}}{\sqrt{N}}$, $\omega_{\mathrm{o},n}\!\approx\!\omega_{\mathrm{o}}\pm(n-1)\tfrac{\Delta\omega}{N}$. The negative sign corresponds to $\theta\!<\!45^{\circ}$, whereupon $\omega_{\mathrm{o},N}\!=\!\omega_{\mathrm{o}}$; and the positive sign used otherwise, whereupon $\omega_{\mathrm{o},1}\!=\!\omega_{\mathrm{o}}$; examples of the latter case are depicted in Fig.~\ref{fig:Concept}(b,c). 

It is clear from comparing Fig.~\ref{fig:Concept}(a)-iv with Fig.~\ref{fig:Concept}(c)-iv that spectral recycling results in added structure to the spatio-temporal intensity profile. To gain insight into these changes, we write the spectrally recycled ST wave packet as 
\begin{equation}
E_{\mathrm{rec}}(x,z;t)=\sum_{n=1}^{N}e^{i(k_{\mathrm{o},n}z-\omega_{\mathrm{o},n}t)}\psi_{n}(x,z;t),
\end{equation}
where $\psi_{n}(x,z;t)$ is the envelope associated with the $n^{\mathrm{th}}$ sub-bandwidth,
\begin{equation}
\psi_{n}(x,z;t)=\int_{\Delta\omega_{n}}\!\!\!dk_{x}\,\widetilde{\psi}_{n}(k_{x})e^{ik_{x}x}e^{-i(\omega(k_{x})-\omega_{\mathrm{o},n})(t-z/\widetilde{v})}.
\end{equation}
Note that $\psi_{n}(x,z;t)\!=\!\psi_{n}(x,0;t-z/\widetilde{v})$, so that the envelope associated with each sub-bandwidth also propagates rigidly at the same group velocity $\widetilde{v}\!=\!c\tan{\theta}$ as its non-recycled counterpart associated with the full temporal bandwidth $\Delta\omega$.

A particularly simple case is that of a narrowband wave packet whereupon $\Delta\omega_{n}\!\approx\!\tfrac{\Delta\omega}{N}$, and the associated envelopes $\psi_{n}(x,z;t)$ are thus almost identical, $\psi_{n}(x,z;t)\approx\psi_{1}(x,z;y)$ for all $n$, in which case:
\begin{equation}\label{eq:RecycledProfile}
I_{\mathrm{rec}}(x,z;t)\propto I_{1}(x,z;t)I_{\mathrm{f}}(z-ct),
\end{equation}
where $I_{1}(x,z;t)\!=\!|\psi_{1}(x.z;t)|^{2}$ and the form factor $I_{\mathrm{f}}(z-ct)$ is the familiar term resulting from a phasor sum,
\begin{equation}\label{eq:FormFactor}
I_{\mathrm{f}}(z-ct)=\frac{\sin^{2}{\left(\frac{\Delta\omega_{1}}{2}N(\frac{z}{c}-t)\right)}}{\sin^{2}{\left(\frac{\Delta\omega_{1}}{2}(\frac{z}{c}-t)\right)}}.
\end{equation}
The spectrally recycled ST wave packet is therefore the product of two envelopes: the first is the ST wave packet envelope $I_{1}(x,z;t)$ that corresponds to a single sub-bandwidth and the second is the form factor $I_{\mathrm{f}}(z-ct)$ that is independent of $x$. Temporal walk-off occurs between $I_{1}(x,z;t)$ that travels at $\widetilde{v}$ and $I_{\mathrm{f}}(z-ct)$ that travels at $c$.

\begin{figure}[t!]
\begin{center}
\includegraphics[width=8.6cm]{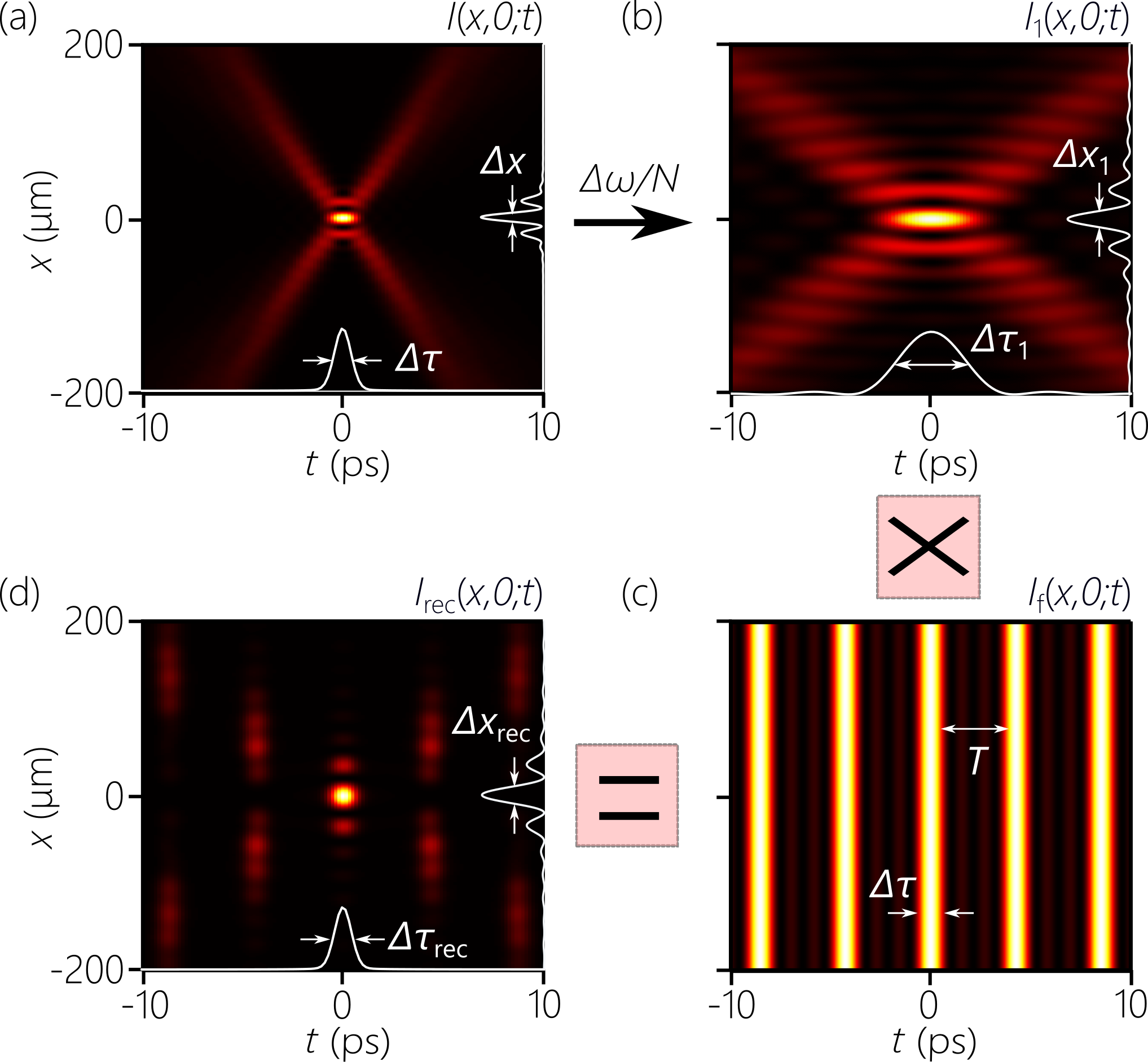} 
\end{center}
\caption{Change in the spatio-temporal profile $I(x,z;t)$ at a fixed axial plane $z\!=\!0$ as a result of spectral recycling. (a) The spatio-temporal intensity profile $I(x,0;t)$ for a non-recycled ST wave packet with spectral tilt angle $\theta\!=\!40^{\circ}$ and temporal bandwidth $\Delta\lambda\!=\!2$~nm. (b) Spatio-temporal intensity profile $I_{1}(x,0;t)$ for a ST wave packet corresponding to a \textit{single} sub-bandwidth $\Delta\omega_{1}$ after spectrally recycling the ST wave packet in (a) with $N\!=\!4$; the temporal bandwidth is $\Delta\lambda/N\!=\!0.5$~nm. (c) The spatio-temporal intensity of the form-factor $I_{\mathrm{f}}(z-ct)$ ensuing from the process of spectral recycling (Eq.~\ref{eq:FormFactor}). (d) The spatio-temporal intensity profile $I_{\mathrm{rec}}(x,0;t)$ for a spectrally recycled ST wave packet with $N\!=\!4$ corresponding to the wave packet in (a), and resulting from multiplying the profiles in (b) and (c). The spatial width has increased $\Delta x_{\mathrm{rec}}\!>\!\Delta x$, but the temporal width remains constant $\Delta\tau\!=\!\Delta\tau_{\mathrm{rec}}$.}
\label{fig:Concept2}
\end{figure}

To elucidate the structure of spectrally recycled ST wave packets, we plot first in Fig.~\ref{fig:Concept2}(a) the spatio-temporal profiles of a non-recycled ST wave packet $I(x,z;t)$ of temporal bandwidth $\Delta\omega$, and the envelope $I_{1}(x,z;t)$ for a single sub-bandwidth $\Delta\omega_{1}\approx\tfrac{\Delta\omega}{N}$ in Fig.~\ref{fig:Concept2}(b), whose pulsewidths at the beam center are $\Delta\tau\!\sim\!\tfrac{2\pi}{\Delta\omega}$ and $\Delta\tau_{1}\!\sim\!\tfrac{2\pi}{\Delta\omega_{1}}\!=\!N\Delta\tau$, respectively. Similarly, the beam widths at the pulse center are $\Delta x\!\sim\!\tfrac{2\pi}{\Delta k_{x}}$ and $\Delta x_{1}\!\sim\!\tfrac{2\pi}{\Delta k_{x}'}\!\sim\!\sqrt{N}\Delta x\!<\!\Delta x$. This is expected because of the smaller spatial and temporal bandwidths associated with $I_{1}(x,z;t)$ in comparison to those associated with $I(x,z;t)$.

Including the $N$ sub-bandwidths in the spectrally recycled ST wave packet results in multiplication of $I_{1}(x,z;t)$ by the form factor $I_{\mathrm{f}}(z-ct)$ plotted in Fig.~\ref{fig:Concept2}(c), resulting in the spatio-temporal intensity profile $I_{\mathrm{rec}}(x,z;t)$ shown in Fig.~\ref{fig:Concept2}(d), where the periodic structure of $I_{\mathrm{f}}(z-ct)$ modulates $I_{\mathrm{rec}}(x,z;t)$. The pulsewidth at the beam center of the spectrally recycled wave packet is $\Delta\tau_{\mathrm{rec}}\!\sim\!\tfrac{4\pi}{N\Delta\omega_{1}}\!=\!\Delta\tau$, similar to that of the non-recycled ST wave packet $I(x,z;t)$. Nevertheless, because $I_{\mathrm{f}}(z-ct)$ is independent of $x$, the beam width at the pulse center remains larger than that for $I(x,z;t)$, $\Delta x_{\mathrm{rec}}\!=\!\Delta x_{1}\!<\!\Delta x$; indeed, $\Delta x_{1}\!\sim\!\sqrt{N}\Delta x$. We have thus retained the narrow temporal width associated with the full temporal bandwidth, while increasing the beam transverse width and thus reducing the associated numerical aperture. Increasing $N$ increases the separation $T\!=\!N\tfrac{2\pi}{\Delta\omega}$ between the discrete features in the spatio-temporal profile, while decreasing the width of these features. 

We can draw several conclusions from this theoretical analysis. First, spectral recycling does \textit{not} affect the critical properties of a ST wave packet, including its propagation invariance and its group velocity $\widetilde{v}$. Second, the temporal linewidth of the pulse at the beam center of the spectrally recycled ST wave packet is similar to that of its non-recycled counterpart, while maintaining a larger beam width at the pulse center. This indicates that the proportionality between the spatial and temporal bandwidths has been modified \textit{without} changing the spectral tilt angle $\theta$. We now proceed to verify experimentally the concept of spectrally recycling ST wave packets and to validate these theoretical predictions concerning their properties. 
  
\begin{figure}[t!]
\begin{center}
\includegraphics[width=7.6cm]{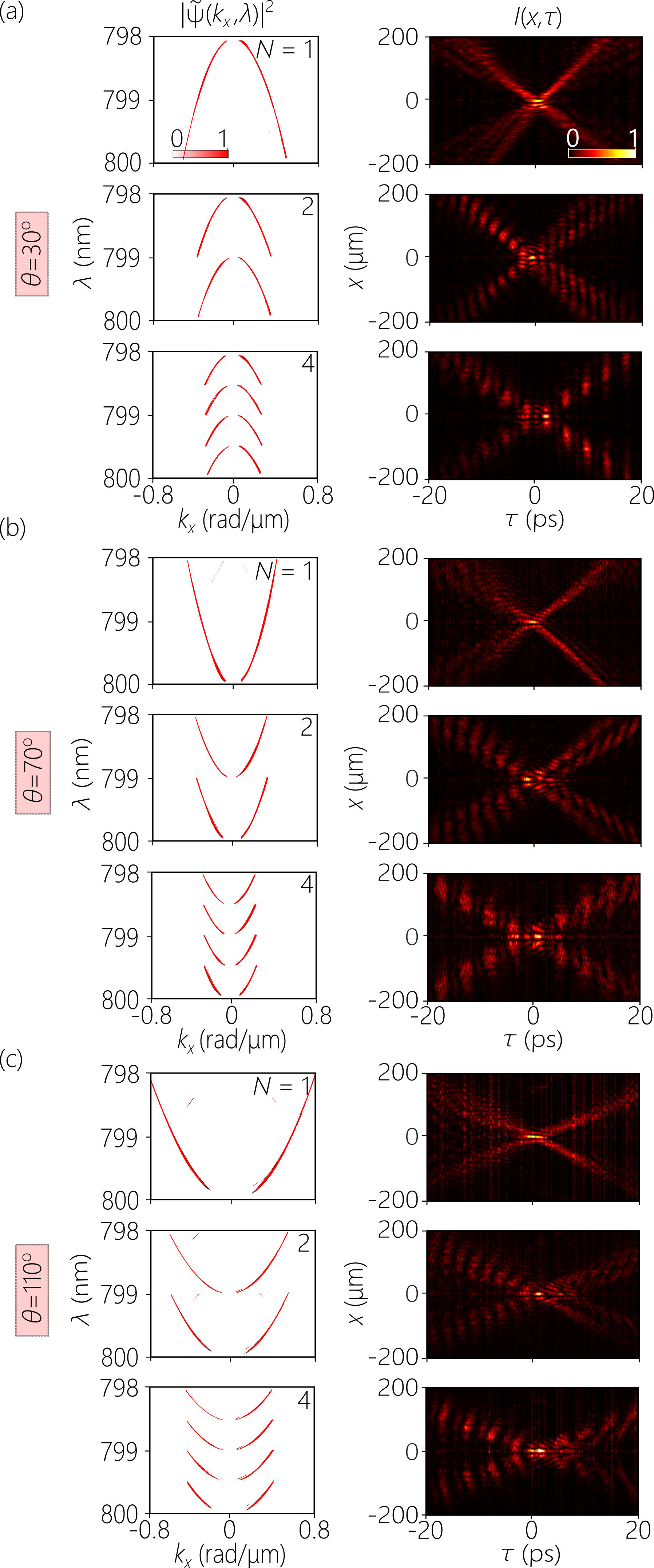} 
\end{center}
\caption{Measurements of the spatio-temporal spectrum $|\widetilde{\psi}(k_{x},\lambda)|^{2}$ (left column) and the spatio-temporal profile $I(x,z;t)$ at a fixed axial plane $z\!=\!10$~mm (right column). (a) The ST wave packet has a spectral tilt angle $30^{\circ}$, (b) $\theta\!=\!70^{\circ}$, and (c) $\theta\!=\!110^{\circ}$. In each panel we plot three cases: $N\!=\!1$ (no spectral recycling), $N\!=\!2$, and $N\!=\!4$.}
\label{fig:Measurements}
\end{figure}

\section{Experimental configuration}

The setup utilized for our proof-of-principle demonstration of spectral recycling is similar to that used previously for synthesizing ST wave packets \cite{Kondakci17NP,Yessenov19PRA,Yessenov19OE}. Beginning with $\sim\!100$-fs pulses from a modelocked Ti:sapphire laser (Tsunami; Spectra Physics) of central wavelength $\sim\!800$~nm, the pulse spectrum is spread via a diffraction grating and collimated by a cylindrical lens before impinging on a two-dimensional, reflective, phase-only spatial light modulator (SLM; Hamamatsu X10468-02). The SLM imparts a phase distribution to the wave front designed to associate each wavelngth $\lambda$ with a prescribed spatial frequency $k_{x}$ in order to satisfy the constraint in Eq.~\ref{eq:OmegaKzEquation}. The retro-reflected field is reconstituted into a ST wave packet at the diffraction grating. We record the spatio-temporal spectrum $|\widetilde{\psi}(k_{x},\lambda)|^{2}$, the spatio-temporal profile of the ST wave packets $I(x,z;\tau)$ at different axial planes $z$, which allows us to assess the group velocity $\widetilde{v}$, in addition to the axial evolution of the time-averaged intensity $I(x,z)$.

We consider three ST wave packets having spectral tilt angles $\theta\!=\!30^{\circ}$ (subluminal $\widetilde{v}\!=\!0.58c$), $\theta\!=\!70^{\circ}$ (superluminal, $\widetilde{v}\!=\!2.75c$) and $\theta\!=\!110^{\circ}$ (negative-$\widetilde{v}$, $\widetilde{v}\!=\!-2.75c$). In each case we carry out the measurements for different extents of spectral recycling: $N\!=\!1$ (non-recycled, $\Delta\lambda\!=\!2$~nm), $N\!=\!2$ ($\Delta\lambda_{n}\approx1$~nm, $n\!=\!1,2$), and $N\!=\!4$ ($\Delta\lambda_{n}\approx0.5$~nm, $n\!=\!1,2,3,4$).

\begin{figure}[t!]
\begin{center}
\includegraphics[width=8.6cm]{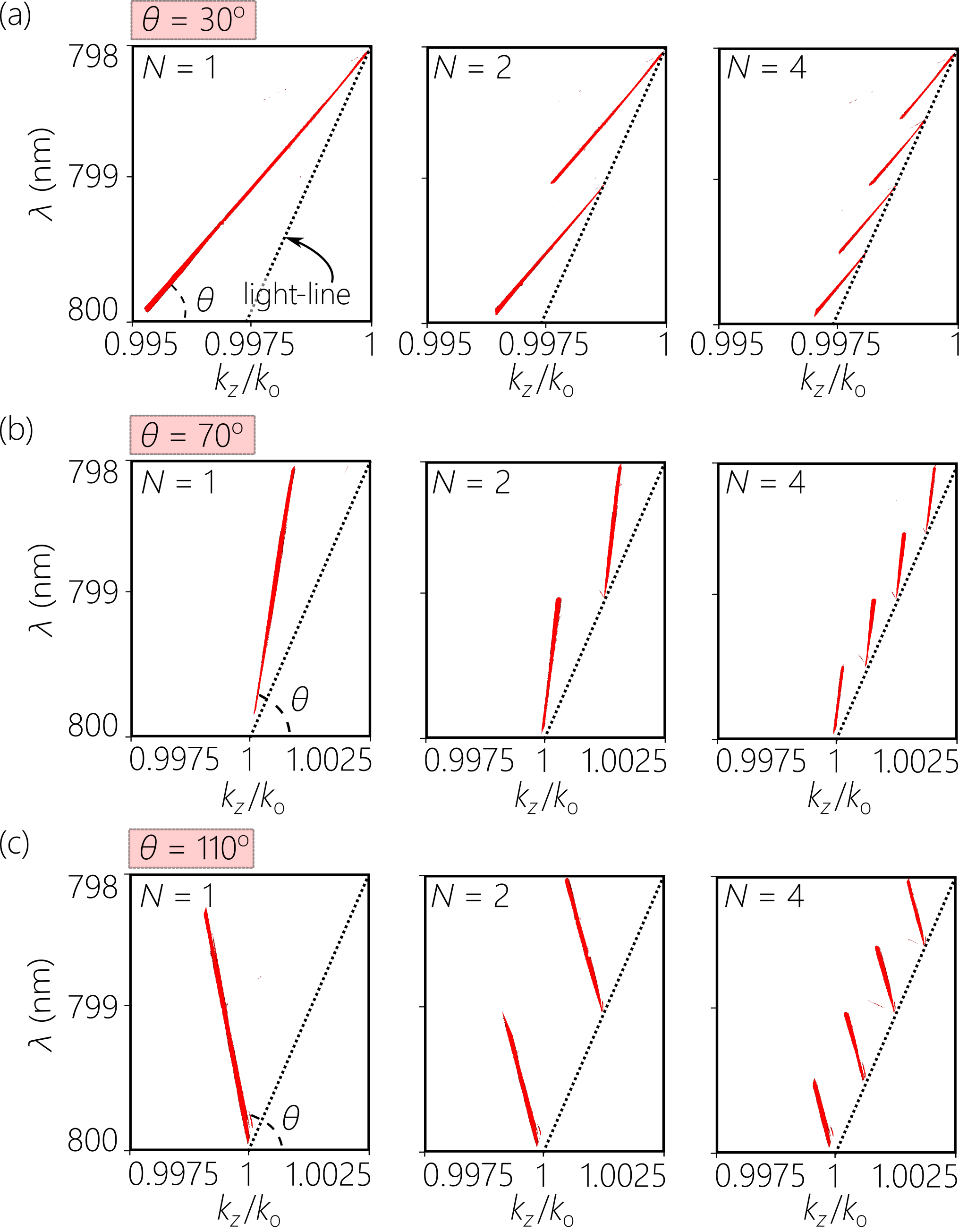} 
\end{center}
\caption{Measurements of the spectral projections onto the $(k_{z},\tfrac{\omega}{c})$-plane for (a) a ST wave packet having a spectral tilt angle $\theta\!=\!30^{\circ}$, (b) $\theta\!=\!70^{\circ}$, and (c) $\theta\!=\!110^{\circ}$. In each panel we plot three cases: $N\!=\!1$ (no spectral recycling), $N\!=\!2$, and $N\!=\!4$, corresponding to the 9 cases shown in Fig.~\ref{fig:Measurements}.}
\label{fig:kzMeasurements}
\end{figure}

\section{Measurements results}

The spatio-temporal spectra for all 9 ST wave packets are plotted in Fig.~\ref{fig:Measurements}. In each case, we observe that the measured spatio-temporal spectrum $|\widetilde{\psi}(k_{x},\lambda)|^{2}$ is divided into $N$ segments all centered at $k_{x}\!=\!0$. Note the switch in sign of the curvature of the spectral projections onto the $(k_{x},\tfrac{\omega}{c})$-plane from the subluminal case ($\theta\!=\!30^{\circ}$) to the superluminal cases ($\theta\!=\!70^{\circ}$ and $110^{\circ}$). The impact of spectral recycling is particularly clear when examining the spectral projection onto the $(k_{z},\tfrac{\omega}{c})$-plane, as shown in Fig.~\ref{fig:kzMeasurements}. Here we see clearly that the spectral projection for the non-recycled ST wave packet extends further away from the light-line (corresponding to larger values of associated $k_{x}$) than in its spectrally recycled counterparts. Spectral recycling divides this projection into $N$ sections, and the segments are shifted accordingly along $k_{z}$ such that the sub-bandwidths are staggered, leaving no spectral gaps. The maximum deviation away from the light-line is reduced as $N$ increases, indicating a reduction in the spatial bandwidth and thus also the numerical aperture. We plot the spatio-temporal profiles $I(x,z;t)$ at fixed $z$ for all these cases in Fig.~\ref{fig:Measurements}. The non-recycled ST wave packets show the expected characteristic X-shape, while those of the spectrally recycled counterparts have the overall X-shaped profile modulated by a discretized structure as predicted in Eq.~\ref{eq:RecycledProfile} and Eq.~\ref{eq:FormFactor}.

\begin{figure}[t!]
\begin{center}
\includegraphics[width=8.6cm]{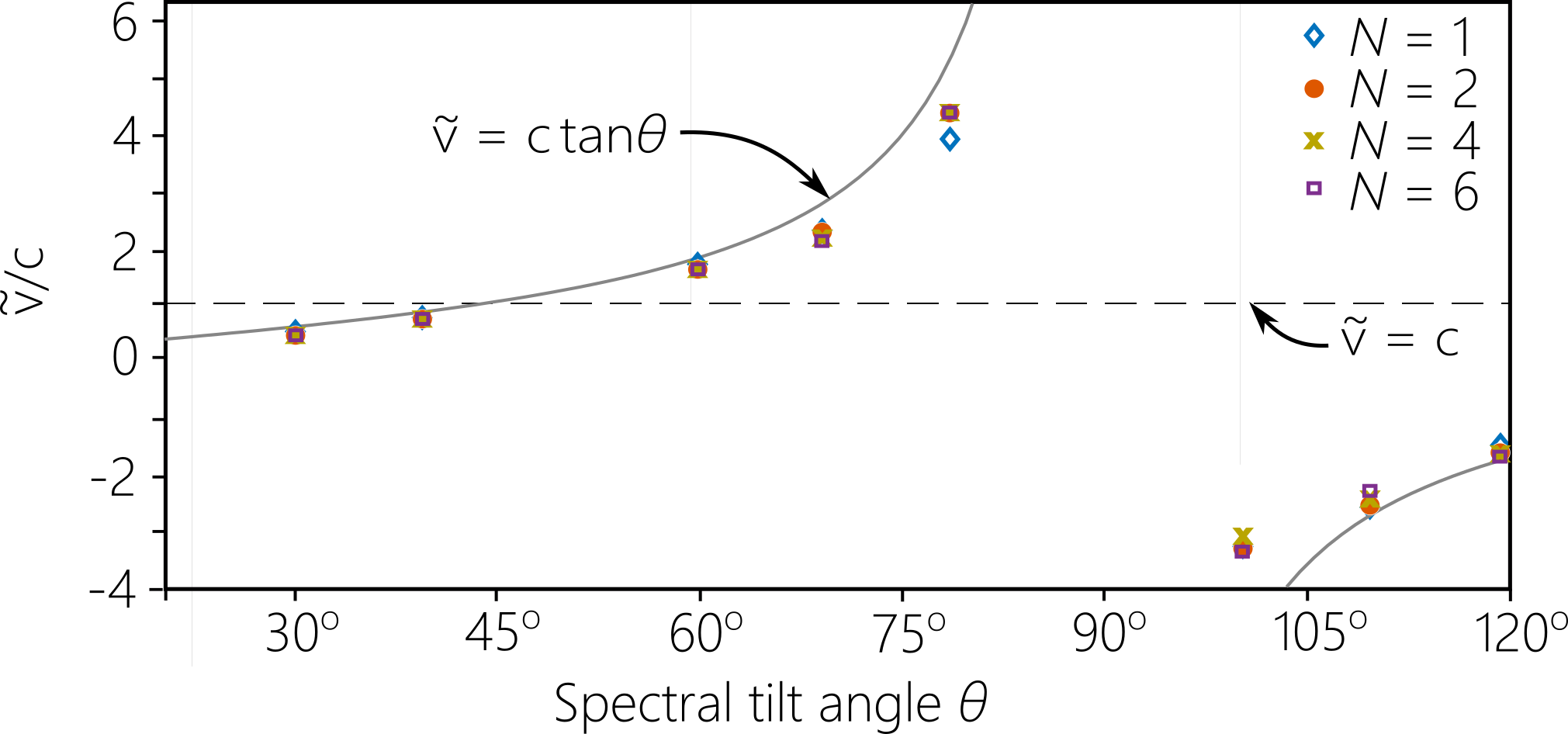} 
\end{center}
\caption{Measurements of the group velocity $\widetilde{v}$ of ST wave packets while varying the spectral tilt angle $\theta$. The solid curve is the theoretical expectation for a non-recycled ST wave packet $\widetilde{v}\!=\!c\tan{\theta}$. The points correspond to measurements for non-recycled ($N\!=\!1$) and spectrally recycled ST wave packets ($N\!=\!2$, 4, and 6). All the data points lie along the theoretical curve, indicating that spectral recycling does not affect $\widetilde{v}$.}
\label{fig:GroupVelocity}
\end{figure}

We next proceed to verify that spectral recycling retains the group velocity of the original non-recycled ST wave packet. The slope of the spectral projection onto the $(k_{z},\tfrac{\omega}{c})$-plane (that is, the spectral tilt angle $\theta$) determines the group velocity. This conclusion has been well-established for non-recycled ST wave packets ($N\!=\!1$) \cite{Kondakci19NC,Yessenov19PRA,Yessenov19OE} and is expected to extend to their spectrally recycled counterparts. To assess the validity of this conception, we measured the group velocity directly using the approach we developed in \cite{Kondakci19NC,Bhaduri19Optica}. This requires measuring the relative group delay between the ST wave packet traveling at $\widetilde{v}$ and a reference laser pulse traveling at $c$, after both wave packets traverse the same distance. From this measured relative group delay we estimate $\widetilde{v}$. In Fig.~\ref{fig:GroupVelocity} we plot the measured $\widetilde{v}$ while varying the spectral tilt angle $\theta$ for different extents of spectral recycling ($N\!=\!1$, 2, 4, and 6). All the measurements follow the same theoretical curve $\widetilde{v}\!=\!c\tan{\theta}$ independently of $N$, thereby confirming that spectral recycling does \textit{not} impact the group velocity of the ST wave packet. 

\begin{figure}[t!]
\begin{center}
\includegraphics[width=8.6cm]{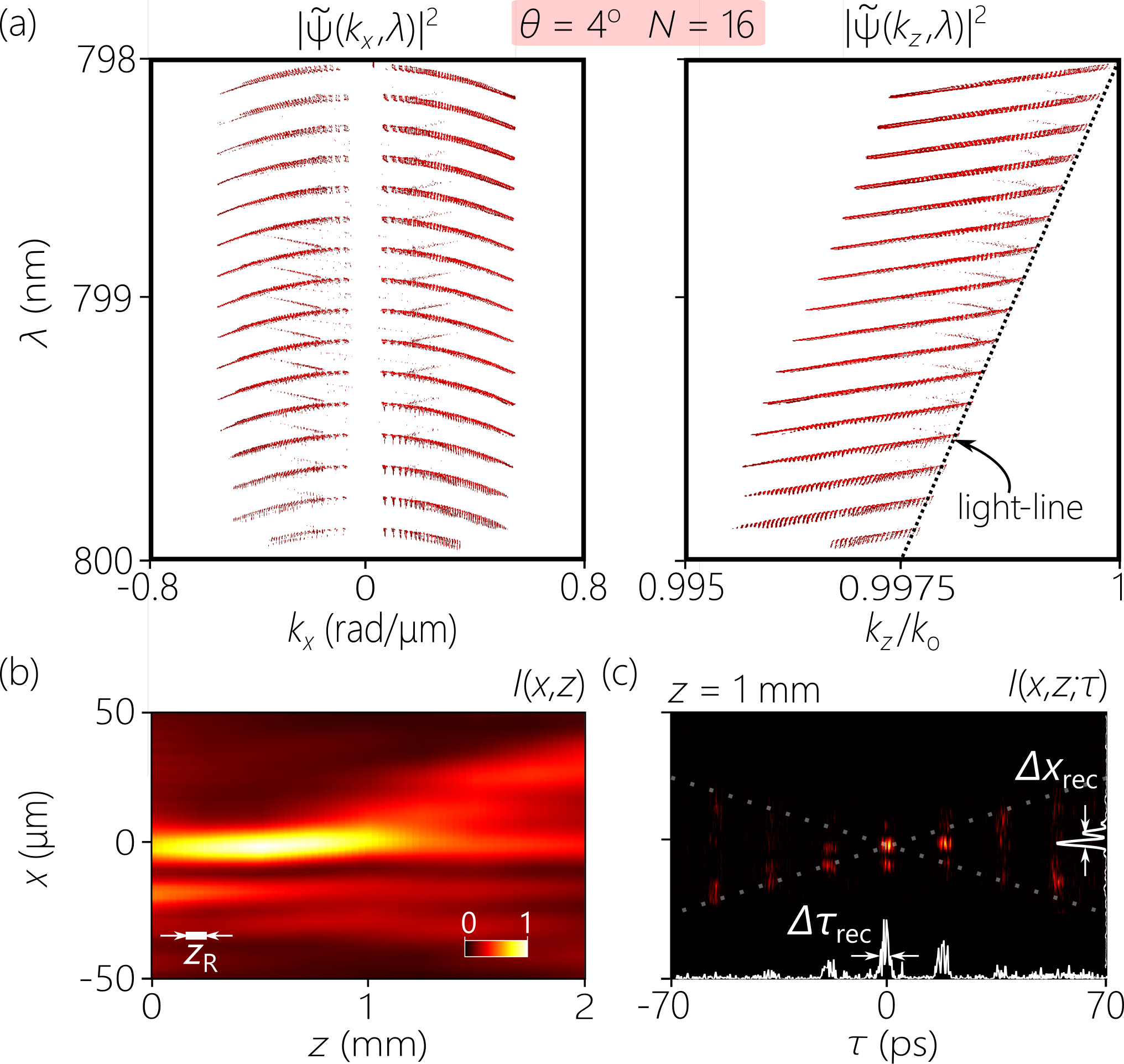} 
\end{center}
\caption{Measurement of an extreme subluminal wave packet at $\theta\!=\!4^{\circ}$ ($\widetilde{v}\!=\!c/14.3$).The ST wave packet is spectrally recycled with $N\!=\!16$. (a) The measured spatio-temporal spectra $|\widetilde{\psi}(k_{x},\lambda)|^{2}$ and $|\widetilde{\psi}(k_{z},\lambda)|^{2}$. (b) The time-averaged intensity $I(x,z)$. (c) The spatio-temporal profile $I(x,z;\tau)$ measured at the axial plane $z\!=\!1$~mm. The measured pulse width at $x\!=\!0$ is $\Delta\tau_{\mathrm{rec}}\!\approx\!3$~ps. The dotted white lines are guides for the eye to discern the X-shape of $I(x,z;\tau)$.}
\label{fig:slow}
\end{figure}

We finally demonstrate that circumventing the usual proportionality between the spatial and temporal bandwidths, $\Delta k_{x}$ and $\Delta\omega$, respectively, through spectral recycling, facilitates for the synthesis of ultraslow ST wave packets. Previously, the slowest subluminal ST we prepared corresponded to $\theta\!=\!10^{\circ}$ ($\widetilde{v}\!\approx\!c/5.7$) \cite{Yessenov19unpub}. In Fig.~\ref{fig:slow} we plot the measurement results for the synthesis of a ST wave packet with $\theta\!=\!4^{\circ}$ ($\widetilde{v}\approx\!c/14.3$). Here, making use of a temporal bandwidth of $\Delta\lambda\!\approx\!2$~nm at $\lambda_{\mathrm{o}}\!=\!800$~nm, synthesizing a ST wave packet with $\theta\!=\!4^{\circ}$ requires a large spatial bandwidth $\Delta k_{x}\!\approx\!0.26k_{\mathrm{o}}$, which is prohibitively difficult to produce. Instead, through spectral recycling with $N\!=\!16$, the spatial bandwidth required is reduced to $\Delta k_{x}\!\approx\!0.064k_{\mathrm{o}}$, which is that corresponding to a non-recycled ST wave packet having $\widetilde{v}\!=\!c/2$ and the same full temporal bandwidth $\Delta\lambda\!\approx\!2$~nm. In Fig.~\ref{fig:slow}(a) we plot the measured spectral projection onto the $(k_{x},\lambda)$ and $(k_{z},\lambda)$ planes, which show clearly the spectrally recycled structure of the spatio-temporal spectrum. We plot in Fig.~\ref{fig:slow}(b) the spatio-temporal intensity profile $I_{\mathrm{rec}}(x,z;t)$. 

\section{Discussion}

A final characteristic of the spectrally recycled ST wave packets to consider is their time-averaged intensity $I(x,z)\!=\!\int\!dt\,I(x,z;t)$, that is captured by a slow detector such as a CCD camera. Because the form-factor $I_{\mathrm{f}}(z-ct)$ is independent of $x$, the time-averaged intensity of the spectrally recycled ST wave packet $I_{\mathrm{rec}}(x,z)$ is diffraction-free with width $\Delta x_{1}$ (corresponding to the spatial bandwidth $\Delta k_{x}'$) of a single sub-bandwidth.

\begin{figure}[t!]
\begin{center}
\includegraphics[width=8.6cm]{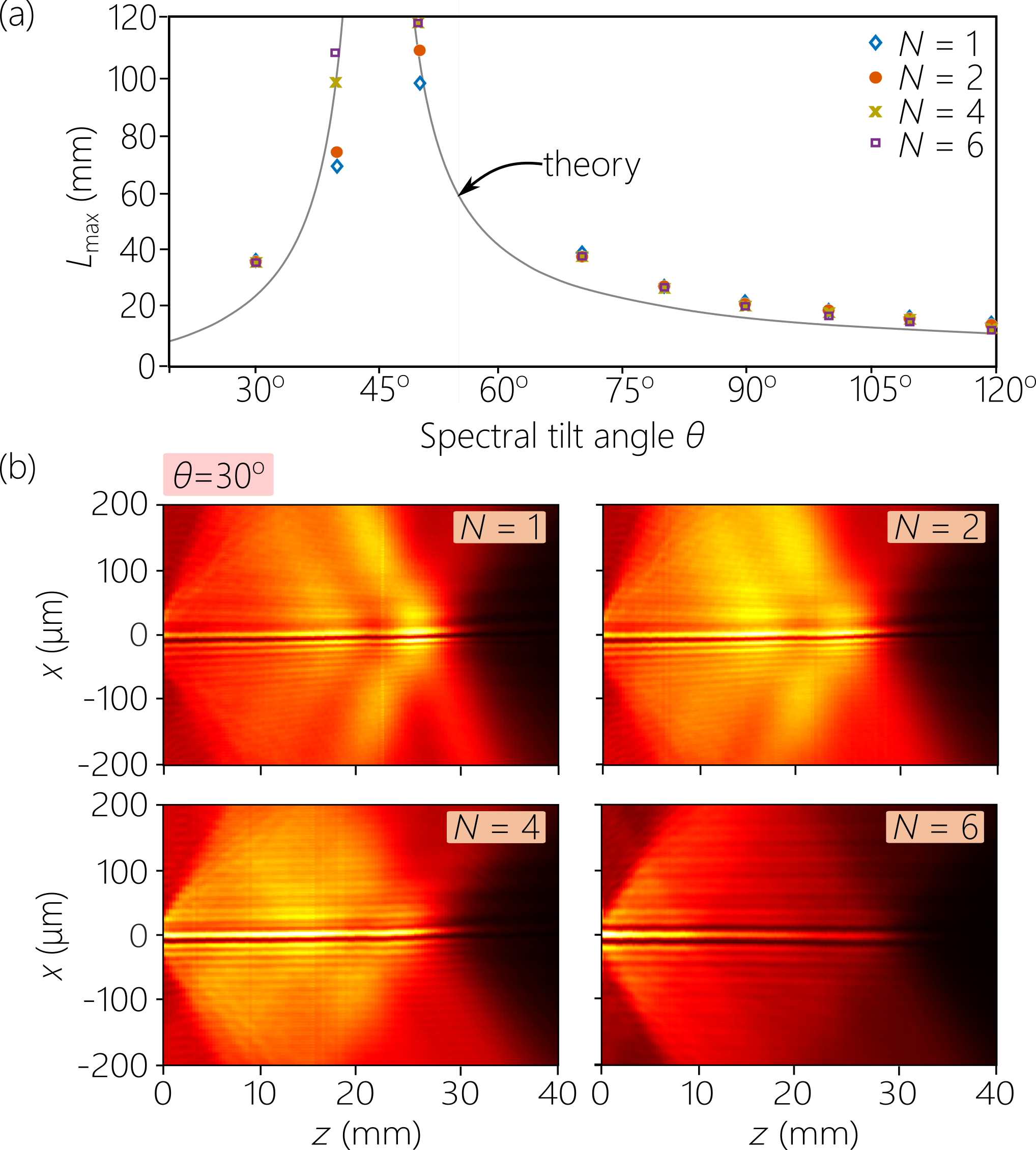} 
\end{center}
\caption{Invariance of the propagation distance $L_{\mathrm{max}}$ with respect to spectral recycling. (a) Measured $L_{\mathrm{max}}$ of spectrally recycled ST wave packet with $N\!=\!1$, 2, 4, and 6. The circles represent data points and the solid curve represents the theoretical expectation for a non-recycled ST wave packet (Eq.~\ref{Eq:LMax}) over a range of spectral tilt angles $\theta\!=\!30^{\circ}\!-\!120^{\circ}$. (b) The axial evolution of the time-averaged intensity $I(x,z)$ of spectrally recycled ST wave packets at $\theta\!=\!30^{\circ}$ for $N\!=\!1$, 2, 4, and 6.}
\label{fig:PropagationDistance}
\end{figure}

The question remains whether spectral recycling affects the maximum propagation distance $L_{\mathrm{max}}$ of a ST wave packet, defined as the propagation distance after which the on-axis intensity drops to half its initial value $I(0,L_{\mathrm{max}})\!=\!0.5I(0,0)$. We have recently shown that $L_{\mathrm{max}}$ for a non-recycled ST wave packet is
\begin{equation}\label{Eq:LMax}
L_{\mathrm{max}}\sim\frac{c}{\delta\omega}\frac{1}{|1-\cot{\theta}|},
\end{equation}
where $\delta\omega$ is the `spectral uncertainty', which is the unavoidable fuzziness in the association between any $k_{x}$ and its assigned $\omega$ \cite{Yessenov19OE}. An ideal delta function correlation in the spatio-temporal spectrum $\widetilde{\psi}(k_{x},\Omega)\!\rightarrow\!\widetilde{\psi}(k_{x})\delta(\Omega-\Omega(k_{x}))$ requires infinite energy and cannot be realized in any finite optical arrangement \cite{Sezginer85JAP,Ziolkowski85JMP}. Instead, in realistic systems \cite{Ziolkowski93JOSAA,Zamboni06JOSAA}, $\widetilde{\psi}(k_{x},\Omega)\!\rightarrow\!\widetilde{\psi}(k_{x})h(\Omega-\Omega(k_{x}))$, where $h(\Omega)$ is a narrow spectral function of width $\delta\omega$, which then determines $L_{\mathrm{max}}$ \cite{Yessenov19OE,Kondakci19OL}. Counter-intuitively, $L_{\mathrm{max}}$ is \textit{not} related to the relative measure $\tfrac{\Delta\omega}{\delta\omega}$ between the full bandwidth $\Delta\omega$ and the spectral uncertainty $\delta\omega$. Instead, $L_{\mathrm{max}}$ is dictated by the absolute value of $\delta\omega$ independently of $\Delta\omega$. This suggests that $L_{\mathrm{max}}$ is also unaffected by spectral recycling as long as the recycling process does not impact $\delta\omega$. Measurements support this suggestion as shown in Fig.~\ref{fig:PropagationDistance}(a) where we plot $L_{\mathrm{max}}$ while varying $\theta$ for ST wave packets of different levels of spectral recycling $N$. The measurements all lie along the theoretical expectation for $L_{\mathrm{max}}$ in Eq.~\ref{Eq:LMax} for non-recycled and recycled ST wave packets independently of $N$. This is further highlighted in Fig.~\ref{fig:PropagationDistance}(b) where we plot $I_{\mathrm{rec}}(x,z)$ for a wave packet with $\theta\!=\!30^{\circ}$ ($\widetilde{v}\!\approx\!0.58c$) for $N\!=\!1$, 2, 4, and 6. The propagation distance $L_{\mathrm{max}}$ remains constant despite the increase in transverse width $\Delta x_{1}$ with $N$. Furthermore, the measured time-averaged intensity $I_{\mathrm{rec}}(x,z)$ for the ultraslow ST wave packet in Fig.~\ref{fig:slow}(c) also satisfies the theoretical limit. These results confirm that spectral recycling does not impact $L_{\mathrm{max}}$ negatively.

\section{Conclusions}

In conclusion, we have proposed and demonstrated a technique we have called `spectral recycling' that helps circumvent the proportionality between spatial and temporal bandwidths intrinsic to ST wave packets. In contrast to traditional pulsed optical beams where the spatial and temporal degrees of freedom are more-or-less independent, each spatial frequency $k_{x}$ in a ST wave packet is typically associated with a single temporal frequency $\omega$ (or wavelength). The spectral recycling strategy `reuses' or `recycles' each spatial frequency $k_{x}$ by associating it with $N$ distinct and well-separated temporal frequencies $\omega_{1},\omega_{2},\cdots,\omega_{N}$ to maintain a fixed spatial bandwidth independently of an increasing temporal bandwidth. We confirmed experimentally that spectral recycling even up to $N\!=\!16$ does \textit{not} affect the unique characteristics of a ST wave packet: its propagation invariance, its group velocity $\widetilde{v}$, and its maximum propagation distance $L_{\mathrm{max}}$ all remain intact. This strategy facilitates the synthesis of ultraslow ST wave packets in free space without exorbitant increase in the required numerical aperture. Indeed, we demonstrated here $\widetilde{v}\!=\!c/14.3$ with the same numerical aperture associated with $\widetilde{v}\!=\!c/2$. It is expected that spectral recycling will enable the utilization of broadband optical sources in the synthesis of ST wave packets, while maintaining a low numerical aperture. This will be particularly useful in application in which a broadband ST wave packet is required, such as in omni-resonant coupling to planar cavities \cite{Shabahang17SR,Shiri20OL}, nonlinear optics, and laser-plasma interactions. Finally, it would also be interesting to examine whether spectral recycling can benefit another recently developed class of optical wave packets having controllable group velocity known as a `flying focus' \cite{SaintMarie17Optica,Froula18NP,Jolly20OE}.

\section*{Acknowledgments}
This work was funded by the U.S. Office of Naval Research (ONR) under contract N00014-17-1-2458, and ONR MURI contract N00014-20-1-2789.

\bibliography{diffraction}

\end{document}